\documentclass[prb,aps,amsmath,preprint]{revtex4}
\usepackage{graphicx}
\usepackage[version=3]{mhchem}
\usepackage{dcolumn}
\usepackage{multirow}
\usepackage{bm}
\usepackage{subfigure}
\usepackage{amssymb}

\begin{document}

\title{Pressure induced structural phase transition in solid oxidizer KClO$_3$:  A first-principles study}
\author{N. Yedukondalu, Vikas D. Ghule, and G. Vaitheeswaran$^*$ }
\affiliation{Advanced Centre of Research in High Energy Materials (ACRHEM),
University of Hyderabad, Prof. C. R. Rao Road, Gachibowli, Andhra Pradesh, Hyderabad- 500 046, India.}
\date{\today}

\begin{abstract}
High pressure behavior of potassium chlorate (KClO$_3$) has been investigated from 0-10 GPa by means of first principles density functional theory (DFT) calculations. The calculated ground state parameters, transition pressure and phonon frequencies using semiempirical dispersion correction scheme are in excellent agreement with experiment. It is found that KClO$_3$ undergoes a pressure induced first order phase transition with an associated volume collapse of 6.4$\%$ from monoclinic (\emph{P2$_1$/m}) $\rightarrow$ rhombohedral (\emph{R3m}) structure at 2.26 GPa, which is in good accord with experimental observation. However, the transition pressure was found to underestimate (0.11 GPa) and overestimate (3.57 GPa) using LDA and GGA functionals, respectively. Mechanical stability of both the phases are explained from the calculated single crystal elastic constants. In addition, the zone center phonon frequencies have been calculated using density functional perturbation theory at ambient as well as at high pressure and the lattice modes are found to soften under pressure between 0.6 to 1.2 GPa. The present study reveals that the observed structural phase transition leads to changes in the decomposition mechanism of KClO$_3$ which corroborates with the experimental results.  
\end{abstract}

\maketitle
\section {Introduction}
Explosives usually contain a mixture of fuels and oxidizers in appropriate proportions to enhance combustion and to release large amount of energy at the expense of chemical reactions occurring in the system. Oxidizers such as metal or ammonium-based nitrates, chlorates, perchlorates, permanganates, chromates, peroxides are frequently used in the explosives and pyrotechnic systems.\cite{steinhauser,kubota} Explosive materials undergo decomposition and lead to high energy release via oxidation process. Most of the secondary explosives have negative oxygen balance (oxygen deficiency) and their heat release is restricted due to incomplete oxidation reactions.\cite{akhavan} Hence, oxidizers are mixed with energetic materials to make the oxidation reactions more complete and reduce the demand of oxygen from the air.\cite{hemmila,conkling,berger,dong} It is a well-known fact that oxidizer composites respond easily to an external stimulus such as impact or friction due to their sensitivity to mechanical energies.\cite{takita,conkling1} The explosive properties of chlorate-based energetic compositions were reported at the end of the 18$^{th}$ century on mixtures of potassium chlorate with sugar and sulfur.\cite{comet,oxley} KClO$_3$ would be an ideal choice as oxidizer because of its ability to sustain at low reaction temperatures, kinetic stability at its melting point, and compatibility with energetic materials.\cite{dolata,markowitz, urban} It is more appropriate to use KClO$_3$ in pyrotechnic mixtures due to its lower energy content and the combustion behavior. Dong and Liao et al\cite{dong,liao} investigated the thermal decomposition process of the KClO$_3$-RDX and KClO$_3$-HMX composite materials and proved that energy release of mixtures exceeded that of pure RDX and HMX, respectively by 10\% and 40\%. In addition, their study also revealed that the presence of KClO$_3$ increases the gaseous products in the decomposition mechanism. The decomposition process of KClO$_3$ takes place in two different mechanisms.\cite{dong,wendlandt,otto,johnson} 
\begin{equation}
2KClO_3 (Monoclinic) \rightarrow 2KCl (B1) + 3O_2 \\
\end{equation}
\begin{equation}
4KClO_3 (Rhombohedral) \rightarrow 3KClO_4 (Orthorhombic) + KCl (B2) 
\end{equation}
At high pressures, the first mechanism is almost inoperative, but the second mechanism is actually favored, since the mixture of the decomposition products is denser than KClO$_3$ itself. This process is relatively simple and the reaction mechanism is also easy to obtain. KClO$_3$ provides a large amount of free oxygen radicals in its thermolysis process which is used to oxidize the energetic intermediates produced by thermal decomposition of explosives. 
\par KClO$_3$ is an ionic molecular solid with a positive oxygen balance of 39.2\%,\cite{meyer} which crystallizes in the primitive monoclinic structure having space group \emph{P2$_1$/m} with lattice parameters a = 4.6569$\AA$, b = 5.59089$\AA$, c = 7.0991$\AA$, $\beta$ = 109.648$^\circ$ and Z = 2 at ambient conditions.\cite{danielsen} The electronic structure and optical properties of KClO$_3$ were reported by Zhuravlev et al\cite{zhuravlev} and Vorobeva et al,\cite{vorobeva} respectively. High pressure studies on molecular crystals are interesting because the effect of pressure on intermolecular bonds is predominant over intramolecular bonds.\cite{sherman} So far, several Raman and IR measurements have been performed on KClO$_3$ at ambient as well as at extreme conditions.\cite{danielsen,sherman,bates,kumari,ramdas,heyns,brooker,adams,bridgman} The Raman studies reveal that KClO$_3$ undergoes a pressure induced structural phase transition from ambient monoclinic (phase I) to rhombohedral (phase II) structure about 0.7 GPa,\cite{heyns} or above 1 GPa.\cite{pistorius1,pistorius2} Moreover, recent high pressure X-ray diffraction study (XRD)\cite{pravica1,pravica2} shows that KClO$_3$ undergoes a rapid decomposition below 2 GPa, and it slowdown above 2 GPa because of structural phase transformation from phase I to II. Fundamental properties such as structural and dynamical stability are still not fully understood due to lack of theoretical data at ambient as well as at high pressures for KClO$_3$. Further, these properties will be of help in understanding the decomposition mechanisms of KClO$_3$. Therefore, in the present study, we have systematically investigated the effect of pressure on crystal structure and lattice dynamics of KClO$_3$ by means of the first principles calculations based on the density functional theory (DFT). The remainder of the article is organized as follows: in section II, we briefly describe the methodology of our calculation. In section III, the structural and vibrational properties of KClO$_3$ at ambient as well as at high pressure are discussed. Finally, in section IV, we summarize the results.

\section{Computational details}
CAmbridge Series of Total Energy Package (CASTEP)\cite{Segall} has been used to perform the first-principles calculations of KClO$_3$ using plane wave  pseudopotential (PW-PP) approach based on DFT.\cite{Hohenberg, Kohn} We used Norm-conserving pseudopotentials\cite{Troullier} for electron-ion interactions. The local density approximation (LDA)\cite{Ceperley,Perdew} and generalized gradient approximation (GGA)\cite{Burke} were used to treat electron-electron interactions in KClO$_3$. The Broyden-Fletcher-Goldfarb-Shanno (BFGS) minimization scheme\cite{Almlof} has been used for structural relaxation. The plane wave basis orbitals used in the calculations are 3$s^2$, 3$p^6$, 4$s^1$ for K; 3$s^2$, 3$p^5$ for Cl and 2$s^2$, 2$p^4$ for O. The convergence criteria for structural optimization for both phases I and II of KClO$_3$ was set to ultrafine quality with a kinetic energy cutoff of 1000 eV and 2$\pi\times$0.04 $\AA^{-1}$ separation of k-mesh according to the Monkhorst-Pack grid scheme.\cite{Monkhorst} The self-consistent energy convergence was set to 5.0$\times$10$^{-6}$ eV/atom. The convergence criterion for the maximum force between atoms was 0.01 eV/A. The maximum displacement and stress were set to be 5.0$\times$10$^{-4}\AA$ and 0.02 GPa, respectively. The standard LDA (CA-PZ) and GGA (PBE) functionals are inadequate to predict the long range interactions in molecular crystalline solids.\cite{Santra, Lu, Dion} However, the weak van der Waals interactions play a key role in determining the physical and chemical properties of molecular solids. In order to account for the long range interactions in the molecular solids, semiempirical dispersion correction schemes have been developed and incorporated in the standard DFT description. In the present study, we have used G06 scheme\cite{Grimme} implemented through the GGA functional (PBE+G06), which is an empirical correction to DFT taking into account the dispersive interactions based on damped and atomic pairwise potentials of the form C$_6$$\times$R$^{-6}$. We have systematically studied the effect of semiempirical dispersion correction scheme on the structural and vibrational properties of the KClO$_3$ at ambient as well as at high pressure along with the standard DFT functionals, which are discussed in detail in the following sections.

\section{Results and discussion}
\subsection{Structural properties at ambient pressure}
XRD studies on single crystals of KClO$_3$ reveal that the phase I of KClO$_3$ possesses distorted NaCl-type structure with low symmetry, crystallizing in the primitive monoclinic crystal structure with \emph{P2$_1$/m} space group. By taking experimental crystal structure as input,\cite{danielsen} we performed full structural optimization of the KClO$_3$ including lattice parameters and internal coordinates using standard LDA and GGA functionals. It is found that the calculated volume is underestimated by 12.6$\%$ within CA-PZ and overestimated by 7.5$\%$ within PBE functionals when compared to experimental volume and thus, the PBE volume is relatively closer to experiment than CA-PZ. As discussed in section I, the phase I of KClO$_3$ has two molecules per unit cell, which are held together through the weak van der Waals interactions in the crystal structure, which leads to the large discrepancy between the calculated ground state parameters and the experimental data. Thus the standard exchange-correlation potentials used in the calculations do not capture the nature of non-bonded interactions in such molecular solids. Therefore, we performed structural relaxation with the semiempirical dispersion correction scheme PBE+G06, to account for the non-bonded interactions in our calculations. The optimized equilibrium crystal structure of phase I of KClO$_3$ is shown Fig 1a. The calculated lattice constants, and volume with the standard DFT functionals and semiempirical dispersion correction scheme are summarized in Table I and the fractional co-ordinates are given in Table I (see supporting information). The obtained equilibrium lattice constants and volume using the PBE+G06 are in very good agreement with the experiment\cite{danielsen} and the corresponding CA-PZ and PBE values show large deviation from the experimental results at ambient pressure.  

\begin{figure}[h]
\centering
\includegraphics[height = 2.2in, width=6.0in]{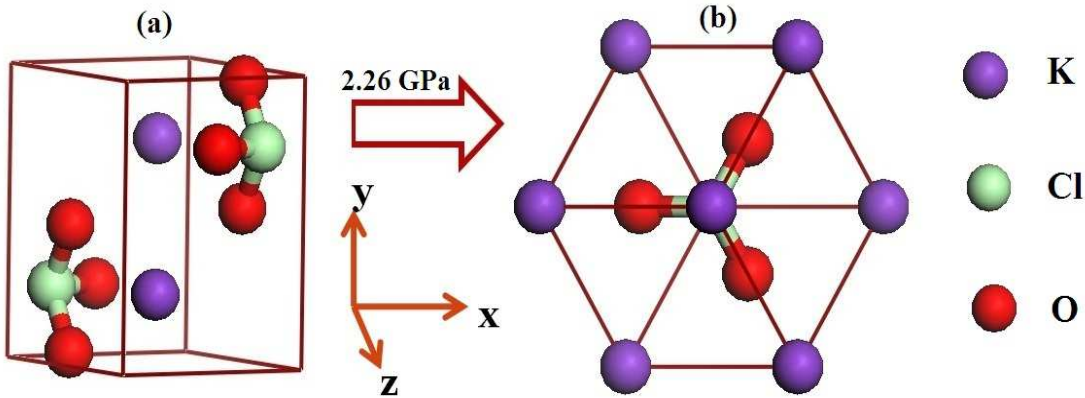}
\caption{(Colour online) Crystal structures of a) phase I (monoclinic, \emph{P2$_1$/m}) and b) phase II (rhombohedral, \emph{R3m}) of KClO$_3$.}
\end{figure}

\subsection{Monoclinic $\rightarrow$ Rhombohedral structural phase transition under pressure}          
The investigation of energetic materials at high pressure are challenging because they possess complex chemical behavior and there is a risk of decomposition. However, theoretical modeling and simulations are efficient tools to predict physical and chemical properties of complex energetic solids at high pressure as a complement to experiments in modern physics, chemistry and materials science. Raman spectroscopic measurements and high pressure XRD studies reveal that KClO$_3$ undergoes a pressure induced phase transition from phase I $\rightarrow$ II about 0.7 GPa,\cite{heyns} 1.0 GPa\cite{pistorius1,pistorius2} and above 2.0 GPa.\cite{pravica2} There is an inconsistency between the experiments\cite{heyns,pistorius1,pistorius2,pravica2} with regard to the transition pressure of KClO$_3$. In order to resolve this issue, first principles total energy calculations were performed based on DFT, which is a powerful tool in predicting the behavior of solid state systems at ambient as well as at high pressures. As illustrated in Fig. 2, the calculated enthalpy difference shows that phase I transforms to phase II about at 0.11 and 3.57 GPa using CA-PZ and PBE, respectively. It is well known from the literature\cite{oganov1,oganov2,kang,jaffe} that the CA-PZ functional usually underestimates transition pressures, whereas the GGA overestimates the same. The same trend is observed in our present calculations for KClO$_3$, where the CA-PZ functional severely underestimates the transition pressure, whereas the PBE functional overestimates when compared to the experimental transition pressure \emph{i.e.}, above 2 GPa.\cite{pravica2} However, the van der Waals corrected PBE+G06 functional provides reliable transition pressures for molecular solids.\cite{griffiths1,griffiths2,kambara} In the present study, the PBE+G06 functional reproduces the transition pressure accurately in contrast to the standard CA-PZ and PBE functionals, and this in very good agreement with the experimental transition pressure.\cite{pravica2} 

\par The calculated lattice parameters a, b, and c of phase I and lattice parameter a of phase II are shown in Fig. 3a as a function of pressure. It is clearly observed that all lattice parameters in both phases decrease monotonically with increasing pressure. But, the angles of both phases ($\beta$ in phase I and $\alpha$ in phase II) increase monotonically over the studied pressure range. Fig. 3b shows the pressure dependence of volume and there is a 6.4\% volume collapse at 2.26 GPa. This is consistent with experimental results as the ambient phase transforms to a $\sim$6$\%$ denser phase II\cite{pistorius1,pistorius2} of KClO$_3$, which indicates the first order nature of phase transition. We also predicted the cell parameters of phase II at 3 GPa, which crystallizes in the rhombohedral structure with \emph{R3m} (Z = 1) space group (see Fig. 1b). The calculated lattice parameters a = 4.1789$\AA$ and $\alpha$ = 84.95$^\circ$ are in good agreement with experimental data at $\sim$3.5 GPa, a = 4.201$\AA$ and $\alpha$ = 84.80$^\circ$.\cite{pistorius2,pravica2} All structural parameters of phase II are presented in Table II along with experimental data.\cite{pistorius2} In addition, the pressure-volume data of phase II are consistent with high pressure XRD data.\cite{pravica2} 

\begin{figure}[h]
\centering
\includegraphics[height = 3.5in, width=4.0in]{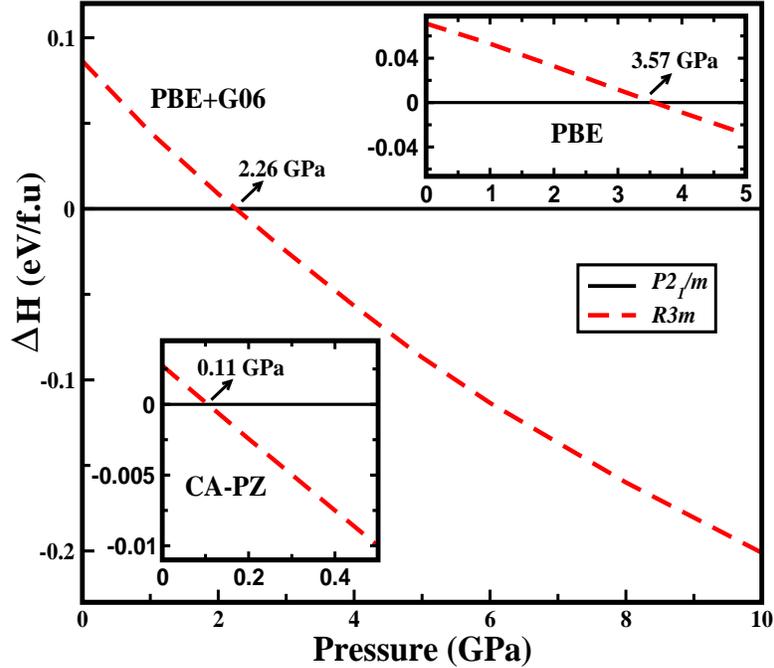}
\caption{(Colour online) Calculated enthalpy difference of phase I (\emph{P2$_1$/m}, black solid line) and phase II (\emph{R3m}, red dotted line) of KClO$_3$ as a function of pressure using dispersion corrected PBE+G06 functional. The lower and upper inset figures correspond to CA-PZ and PBE functionals, respectively.}
\end{figure}

\begin{figure}[h]
\centering
\includegraphics[height = 4.5in, width=4.0in]{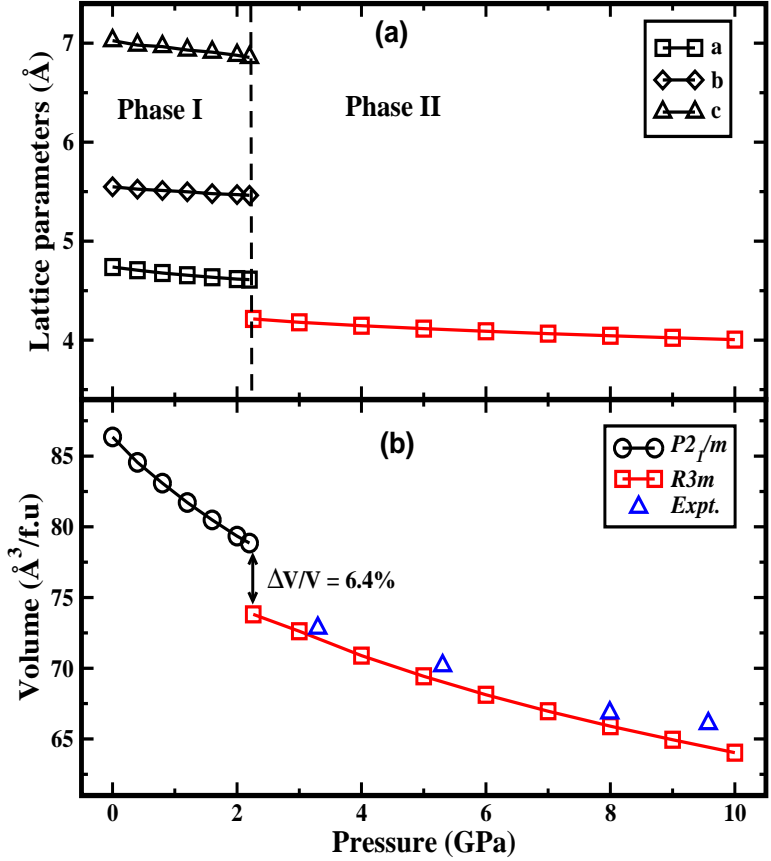}
\caption{(Colour online) Calculated a) Lattice constants b) Volume of phase I and II of KClO$_3$ as a function of pressure using dispersion corrected PBE+G06 functional. The experimental data points are taken from Ref. \onlinecite{pravica2} for the phase II (\emph{R3m}).}
\end{figure}

\par As discussed in an earlier section, the first decomposition mechanism (see section I) of KClO$_3$ is as follows: phase I of KClO$_3$ undergoes a rapid decomposition\cite{pravica1,pravica2} to give B1 form of KCl and oxygen (O$_2$) as the decomposition products at lower pressures ($\textless$ 2 GPa) and this is ineffective at high pressures. The present study reveals that phase I transforms to phase II at 2.26 GPa, which is in good agreement with experimental transition pressure and this structural transition is responsible for the slowdown of the decomposition process of KClO$_3$ above 2 GPa.\cite{pravica1,pravica2}. Since KClO$_3$ transforms from phase I to II  above 2 GPa, the second decomposition mechanism (see section I) is favorable at high pressures. Johnson et al\cite{johnson} reported the detailed thermal decomposition of phase II of KClO$_3$, which produces B2 form of KCl and oxygen (O$_2$) via orthorhombic KClO$_4$. Therefore, one can confirm that the KClO$_3$ can be used as an oxidizer due to release of O$_2$ as one of its decomposition products for oxygen deficient energetic materials like TNT (-74.00\%), HNS (-67.60\%), HMX (-21.62\%), and RDX (-21.60\%).\cite{akhavan}           

\begin{table}[h]
\caption{Calculated lattice parameters (a, b, and c), monoclinic angle ($\beta$), volume (V), density ($\rho$), bond lengths, and bond angles of phase I of KClO$_3$ using LDA, GGA and dispersion corrected PBE+G06 functionals along with experimental data.\cite{danielsen}}
\begin{ruledtabular}
\begin{tabular}{cccccc}
Parameter             &    CA-PZ   &   PBE    &   PBE+G06    &  Expt.\cite{danielsen}   \\ \hline
a($\AA$)              &   4.4617   &  4.7222  &    4.7327    &  4.65690   \\
b($\AA$)              &   5.4149   &  5.7171  &    5.5457    &  5.59089   \\
c($\AA$)              &   6.7850   &  7.2560  &    7.0407    &  7.09910   \\
$\beta(^\circ$)       &   111.81   &  107.22  &    110.94    &  109.648  \\
V($\AA^3$)            &   152.19   &  187.11  &    172.71    &  174.07   \\
$\rho$(g/cc)          &   2.674    &  2.175   &    2.358     &  2.338   \\           
Cl-O1($\AA$)          &   1.481    &  1.498   &    1.501     &  1.499   \\           
Cl-O2($\AA$)          &   1.480    &  1.503   &    1.499     &  1.491   \\           
O1-Cl-O2($^\circ$)    &   106.34   &  106.97  &    106.71    &  106.59  \\           
O2-Cl-O2($^\circ$)    &   106.53   &  106.68  &    106.99    &  106.27  \\           
\end{tabular}
\end{ruledtabular}
\end{table}

\begin{table}[h]
\caption{Calculated lattice parameter (a), rhombohedral angle ($\alpha$), volume (V) and fractional coordinates for phase II of KClO$_3$ at 3.0 GPa using PBE+G06 functional along with the experimental data. \cite{pistorius2,pravica2}}
\begin{ruledtabular}
\begin{tabular}{cccc}
Parameter             & Wyckoff   &   PBE+G06    &      Expt.                         \\ \hline
a($\AA$)              &           &    4.1789    &     4.201$^a$                      \\
$\alpha$ ($^\circ$)   &           &    84.95     &     84.80$^a$                      \\
V($\AA^3$)            &           &    72.17     &     73.28$^a$,77.873$^b$                 \\
K                     &   1a      & (0.0000, 0.0000, 0.0000) &  (0.0000, 0.0000, 0.0000)$^a$ \\
Cl                    &   1a      & (0.4747, 0.4747, 0.4747) &  (0.5000, 0.5000, 0.5000)$^a$ \\
O                     &   3b      & (0.5434, 0.5434, 0.1176) &  (0.6250, 0.6250, 0.1250)$^a$ \\
\end{tabular}
\end{ruledtabular}
$^a$Reference.\cite{pistorius2}, $^b$Reference.\cite{pravica2}
\end{table}

\subsection{Elastic constants and bulk modulus}
The elastic constants are fundamental parameters for crystalline solids which describe stiffness of the solid against externally applied strains. They were calculated using volume-conserving strains technique \cite{Mehl} with the PBE+G06 scheme for both low and high pressure phases of KClO$_3$. A complete asymmetric crystal behavior can be described by 21 independent elastic constants. Due to monoclinic and rhombohedral symmetry, phases I and II of KClO$_3$ possess 13 and 6 independent elastic constants, respectively. The calculated elastic constants for both the phases are presented in Table II (see supporting information) and they obey the Born's mechanical stability criteria\cite{Born} indicating that phase I is mechanically stable at ambient pressure, while the high pressure phase is mechanically stable at 3 GPa. Moreover, the single crystal bulk moduli calculated from the elastic constants for phase I and II are 18.49 GPa and 42.58 GPa, respectively. The calculated bulk modulus value of phase II is in close agreement with the experimentally reported values of 45.1 GPa, \cite{johnson} 42.4 GPa,\cite{adams} and 43.6 GPa.\cite{pravica2} These results reveal that phase I of KClO$_3$ is more compressible compared to phase II.

\subsection{Zone center phonon frequencies at ambient pressure}
A detailed vibrational spectra analysis of phase I of KClO$_3$ has been carried out at ambient conditions as well as at high pressure using linear response method within density functional perturbation theory. Single crystal XRD studies reveal that the phase I of KClO$_3$ belongs to monoclinic \emph{P2$_1$/m} space group with two equivalent ClO$_3^-$ ions located at C$_s$ site. This unit cell consists of 10 atoms, which results in the 30 vibrational modes classified as 3 acoustic and 27 optical modes. According to the group theory analysis of \emph{P2$_1$/m} space group, the group symmetry decomposition into irreducible representations of the modes are as follows:\\
$\Gamma_{acoustic}$ = A$_u$ + 2B$_u$ \\
$\Gamma_{optic}$ = 9A$_g$ + 5A$_u$ + 6B$_g$ + 7B$_u$ \\
All the vibrational modes (A$_g$, B$_g$, A$_u$ and B$_u$) of phase I of KClO$_3$ are summarized as 6A$_u$ + 9B$_u$ + 9A$_g$ + 6B$_g$ due to the centre of symmetry of the crystal structure.\cite{danielsen} A$_g$ and B$_g$ modes possess inversion symmetry and hence they are Raman active, while A$_u$ and B$_u$ modes are IR active due to change of sign under inversion symmetry. We have calculated zone center vibrational frequencies at ambient pressure using CA-PZ, PBE and PBE+G06 functionals (see Tables III and IV). The calculated lattice modes ($\textless$ 200 cm$^{-1}$) using CA-PZ and PBE functionals show large deviation from the experimental results,\cite{bates,heyns,brooker} as the CA-PZ and PBE fuctionals are inadequate to treat intermolecular interactions in molecular solids.\cite{Santra, Lu, Dion} The non-bonded interactions can be treated very well with dispersion corrected PBE+G06 functional\cite{Grimme} and thus, the calculated lattice modes using PBE+G06 functional are found to be in good agreement with the experimental data.\cite{bates,heyns,brooker} The obtained internal (high frequency) modes using CA-PZ, PBE and PBE+G06 functionals are also in close comparison with the experimental results.\cite{bates,heyns,brooker} Further, the optical modes are classified into 15 external ($\textless$ 200 cm$^{-1}$) and 12 internal ($\textgreater$ 400 cm$^{-1}$) modes (see Fig. 4). Among the 15 external (lattice) modes, there are 9 Raman active and 6 IR active modes and these lattice modes are mutually exclusive.\cite{ramdas,kumari} The lattice modes assigned to translational (4A$_g$ + 2B$_u$ + 2B$_g$ + A$_u$) and rotational (A$_g$ + B$_u$ + 2B$_g$ + 2B$_u$) motion of phase I of KClO$_3$ are shown in Table III. 

\begin{table}[h]
\caption{Calculated Raman and IR active lattice modes (in cm$^{-1}$) of phase I of KClO$_3$ at ambient pressure using CA-PZ, PBE and dispersion corrected PBE+G06 functionals along with experimental data.\cite{bates, heyns, brooker} (T = Translational and R = Rotational)} 
\begin{ruledtabular}
\begin{tabular}{cccccccc}
  & Mode & Symmetry  &   CA-PZ   &  PBE   &  PBE+G06  &   Expt.                    & Assignment                  \\ \hline
Raman  & M1  &  A$_g$    &   58.12   & 55.68  &   57.06   &  53$^a$                &    T  \\                          
  & M2  &  B$_g$    &   54.63   & 47.06  &   61.96   &  57,$^a$ 55$^b$             &    T  \\      
  & M3  &  A$_g$    &   90.23   & 78.70  &   87.21   &  86,$^a$ 90,$^b$ 87$^c$     &    T  \\          
  & M4  &  B$_g$    &   86.07   & 66.77  &   97.02   &  78,$^a$ 83$^c$             &    R  \\                  
  & M5  &  A$_g$    &  119.68   & 90.76  &   106.93  &  100,$^a$ 102,$^b$ 109$^c$  &    T  \\        
  & M6  &  B$_g$    &  144.54   & 104.34 &   126.48  &  125,$^a$ 134,$^b$ 132$^c$  &    T  \\
  & M7  &  A$_g$    &  159.80   & 113.55 &   131.80  &     ---                     &    T  \\                                
  & M8  &  B$_g$    &  156.79   & 126.23 &   145.56  &  146,$^b$ 144$^c$           &    R  \\               
  & M9  &  A$_g$    &  179.63   & 130.40 &   148.30  &  145,$^a$ 161$^c$           &    R  \\   
IR&  M10  &  A$_u$   &   83.86   &  49.74  &   88.28   &    ---                    &    R  \\       
  &  M11  &  A$_u$   &   133.24  &  104.73 &  123.34   &    ---                    &    R  \\
  &  M12  &  B$_u$   &   137.77  &  101.53 &  116.02   &    ---                    &    T  \\
  &  M13  &  A$_u$   &   150.69  &  109.31 &  132.73   &   131$^b$                 &    T  \\
  &  M14  &  B$_u$   &   146.85  &  115.35 &  131.17   &    ---                    &    T  \\
  &  M15  &  B$_u$   &   194.60  &  137.83 &  155.52   &    ---                    &    R  \\ 
\end{tabular}
\end{ruledtabular}
$^a$ Reference.\cite{brooker}, $^b$Reference.\cite{heyns}, $^c$Reference.\cite{bates}
\end{table}

\par It is well known that the chlorate ClO$_3^-$ ion in free state possesses four fundamental Raman active vibrations 478 (E), 615 (A$_1$), 930 (A$_1$), and 975 (E) with degeneracies 2, 1, 1, and 2, respectively.\cite{ramdas,kumari} The degeneracy is removed due to the presence of K$^+$ ion in the unit cell of KClO$_3$, which results in six Raman and six IR active non-degenerate internal vibrations. These two sets are mutually exclusive as KClO$_3$ possesses a center of symmetry.\cite{ramdas,kumari} The calculated internal (Raman and IR active) modes using PBE+G06 along with the CA-PZ and PBE functionals and their vibrational assignment are presented in Table IV. The vibrational assignment of internal modes are explained by considering the PBE+G06 modes: the M16 (463.24 cm$^{-1}$) and M22 (463.08 cm$^{-1}$) modes correspond to twisting and deformation of the lattice. The scissoring of Cl-O bond corresponds to M17 (466.45 cm$^{-1}$) and M23 (470.07 cm$^{-1}$) bands, whereas the bending motion of ClO$_3^-$ corresponds to M18 (595.77 cm$^{-1}$) and M24 (594.14 cm$^{-1}$) modes. The modes M19 (916.28 cm$^{-1}$) and M25 (913.00 cm$^{-1}$) correspond to symmetric stretching of Cl-O bonds, while M20 (946.49 cm$^{-1}$), M26 (937.33 cm$^{-1}$) and M21 (953.81 cm$^{-1}$), M27 (927.69 cm$^{-1}$) modes arise due to asymmetric stretching of ClO$_3^-$ ion along yz-plane of the KClO$_3$ lattice. The calculated internal modes with and without dispersion corrected PBE functional are underestimated when compared to the experimental data, which may be due to the fact that the linear response approach used in the present study is based on the harmonic approximation. The calculated internal modes are overestimated within CA-PZ when compared to PBE and PBE+G06 functionals due to underestimation of the unit cell volume and intra molecular bondlengths such as Cl-O1 and Cl-O2 bonds of phase I of KClO$_3$.       

\begin{table}[h]
\caption{Calculated Raman and IR active internal modes (in cm$^{-1}$) of phase I of KClO$_3$ at ambient pressure using CA-PZ, PBE and dispersion corrected PBE+G06 functionals along with experimental data.\cite{bates, heyns, brooker}}
\begin{ruledtabular}
\begin{tabular}{cccccccc}
 & Mode & Symmetry   &   CA-PZ   &  PBE   &  PBE+G06  &   Expt.                & Assignment    \\ \hline
Raman & M16  &  B$_g$     &  477.75   & 456.51 &  463.24   &  487$^a$          &  Twist.+def. of Cl-O      \\                
  & M17  &  A$_g$     &  480.69   & 460.07 &  466.45   &  488,$^a$ 490$^b$     &  Scissor of Cl-O   \\ 
  & M18  &  A$_g$     &  614.84   & 591.79 &  595.77   &  619,$^c$ 620$^{a,b}$ &  Wagg.+Bend. of Cl-O   \\
  & M19  &  A$_g$     &  955.01   & 915.14 &  916.28   &  940,$^{a,b}$ 939$^c$ &  Symm. Str. of Cl-O   \\
  & M20  &  A$_g$     &  987.53   & 952.00 &  946.49   &  979,$^{a,b}$ 978$^c$ &  Asymm. Str. of Cl-O   \\
  & M21  &  B$_g$     &  996.16   & 938.81 &  953.81   &  982,$^{a,c}$ 983$^b$ &  Asymm. Str. of Cl-O   \\ 
IR  & M22  &   A$_u$      &  478.25   &  456.05  &  463.08   &  484$^a$        &  Twist.+def. of Cl-O    \\            
  & M23  &   B$_u$      &  485.02   &  462.73  &  470.07   &  493$^a$          &  Scissor. Cl-O  \\       
  & M24  &   B$_u$      &  614.17   &  590.89  &  594.14   &  620$^a$          &  Wagg.+Bend. of Cl-O  \\       
  & M25  &   B$_u$      &  952.62   &  911.92  &  913.00   &  939$^a$          &  Symm. Str. of Cl-O  \\         
  & M26  &   B$_u$      &  979.78   &  935.45  &  937.33   &  1000$^a$         &  Asymm. Str. of Cl-O  \\      
  & M27  &   A$_u$      &  967.82   &  913.28  &  927.69   &  992$^a$          &  Asymm. Str. of Cl-O  \\         
\end{tabular}
\end{ruledtabular}
$^a$Reference.\cite{bates}, $^b$Reference.\cite{heyns}, $^c$Reference.\cite{brooker}
\end{table}

\subsection{Zone center phonon frequencies under pressure}
\par To investigate the dynamical stability of phase I of KClO$_3$ under hydrostatic pressure, we have systematically studied the effect of pressure on zone center external (lattice) and internal phonon modes using PBE and PBE+G06 functional to complement the high pressure Raman and IR measurements.\cite{heyns}  The calculated phonon density of states (PhDOS) using PBE+G06 at 0.0 and 2.2 GPa are shown in Fig. 4. The low frequency modes (below 200 cm$^{-1}$) are due to both potassium and chlorate ions, whereas modes above 400 cm$^{-1}$ are due to the chlorate anion alone. The modes shif towards a high frequency region around 2.2 GPa. As illustrated in Fig. 5, the low frequency Raman (M1 to M9) and IR (M10 to M15) active lattice modes soften under hydrostatic pressure, except M7 lattice mode which is hardening under pressure. Therefore, the inclusion of dispersion correction (G06) through the PBE functional significantly affects the lattice modes at ambient as well as at high pressure. However, the internal modes are not much affected with the dispersion corrected PBE+G06 functional, which can be clearly seen from Fig. 6. Since, the bond lengths (Cl-O1 and Cl-O2) within the ClO$_3$ molecule are almost unchanged by the inclusion of dispersion correction (PBE+G06) to the standard PBE functional (see Table I), the corresponding ClO$_3$ internal vibrations using PBE and PBE+G06 are found to be comparable with each other (see Table IV and Fig. 6). A similar kind of behavior is observed for energetic molecular crystals such as RDX and HMX at ambient pressure.\cite{vashishta1,vashishta2} The dispersion corrected methods are successful in predicting the lattice vibrations thus enabling comparison to experiments. 
\par The optical modes in the frequency range between 450-620 cm$^{-1}$ i.e., M16-M18 and M22-M24 are found to be slightly modified with the application of pressure, whereas the high frequency internal modes such as M19-M21 and M25-M27 are found to increase linearly with pressure. Overall, we observe that all lattice modes increase with pressure upto 0.6 GPa and remain almost constant between 0.6-0.8 GPa. We also find a sudden decrease in all the Raman and IR active lattice modes from 1.0 to 1.2 GPa, which may results in the dynamical instability of phase I of KClO$_3$ under pressure. As evident from Fig. 5, all the Raman and IR active lattice modes do not vary significantly with pressure after 1.2 GPa, which might be the reason for coexistence of the two phases from 1.0 GPa up to transition pressure (2.26 GPa). Hence, we confirm that the transition starts at about 1.0 GPa due to softening of lattice and completes at 2.26 GPa (see Fig. 2). Thus, low symmetry phase I (\emph{P2$_1$/m}) undergoes a pressure induced structural phase transition to high symmetry phase II (\emph{R3m}), which agrees quite well with the recent experimental results using X-ray diffraction analysis.\cite{pravica2} Further, a detailed lattice dynamical calculations are required to understand the vibrational effects which are responsible for the structural phase transition in KClO$_3$. 

\section{CONCLUSIONS}
In summary, first principles calculations were performed to investigate the high pressure behavior of KClO$_3$. Standard DFT functionals such as LDA and GGA are unable to account for the non-bonded interactions in this molecular solid (see Table I). However, the calculated ground state properties using semiempirical dispersion correction scheme (PBE+G06) are in excellent agreement with experiment. It is found that KClO$_3$ undergoes a pressure induced structural phase transition from phase I to II and the calculated transition pressures using CA-PZ and PBE functionals are 0.11 and 3.57 GPa, respectively. The CA-PZ severely underestimates the transition pressure whereas PBE overestimates it. However, the calculated transition pressure using PBE+G06 functional is 2.26 GPa, which is in good accord with a recent high pressure X-ray diffraction study. The transition is associated with a volume contraction of 6.4$\%$, which indicates a first order type phase transition and this is found to be consistent with experimental reduction of $\sim$6$\%$. The calculated single crystal elastic constants show that phases I and II are mechanically stable. The calculated single crystal bulk modulus (42.58 GPa) and compressibility (2.41$\times$10$^{-2}$ GPa$^{-1}$) of phase II are in excellent agreement with experiments. We also calculated the zone center phonon frequencies at ambient and at high pressures upto 2.2 GPa to investigate the dynamical stability of phase I of KClO$_3$. We observe the softening in the Raman and IR active lattice modes under pressure using PBE+G06 functional. Overall, the present study reveals that the softening of lattice of phase I starts after 1.0 GPa and it completely transforms to phase II at 2.26 GPa, which agrees quite well with recent experimental results. Therefore, the present study confirms that the pressure induced phase transition leads to a change in the decomposition mechanisms of \ce{KClO3} which also supports recent experimental observations.\cite{pravica2} \\ \\

\begin{figure}[h]
\centering
\includegraphics[height = 3.0in, width=2.9in]{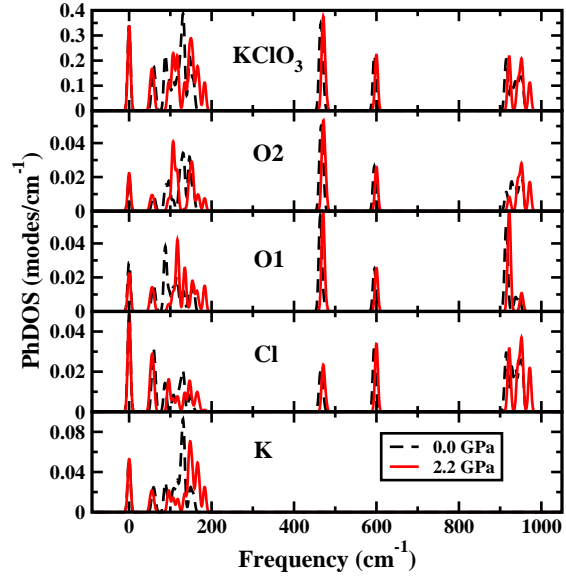}
\caption{(Colour online) Calculated phonon density of states (PhDOS) for phase I of KClO$_3$ using PBE+G06 functional.}
\end{figure}

\begin{figure}[h]
\centering
\includegraphics[height = 3.5in, width=2.9in]{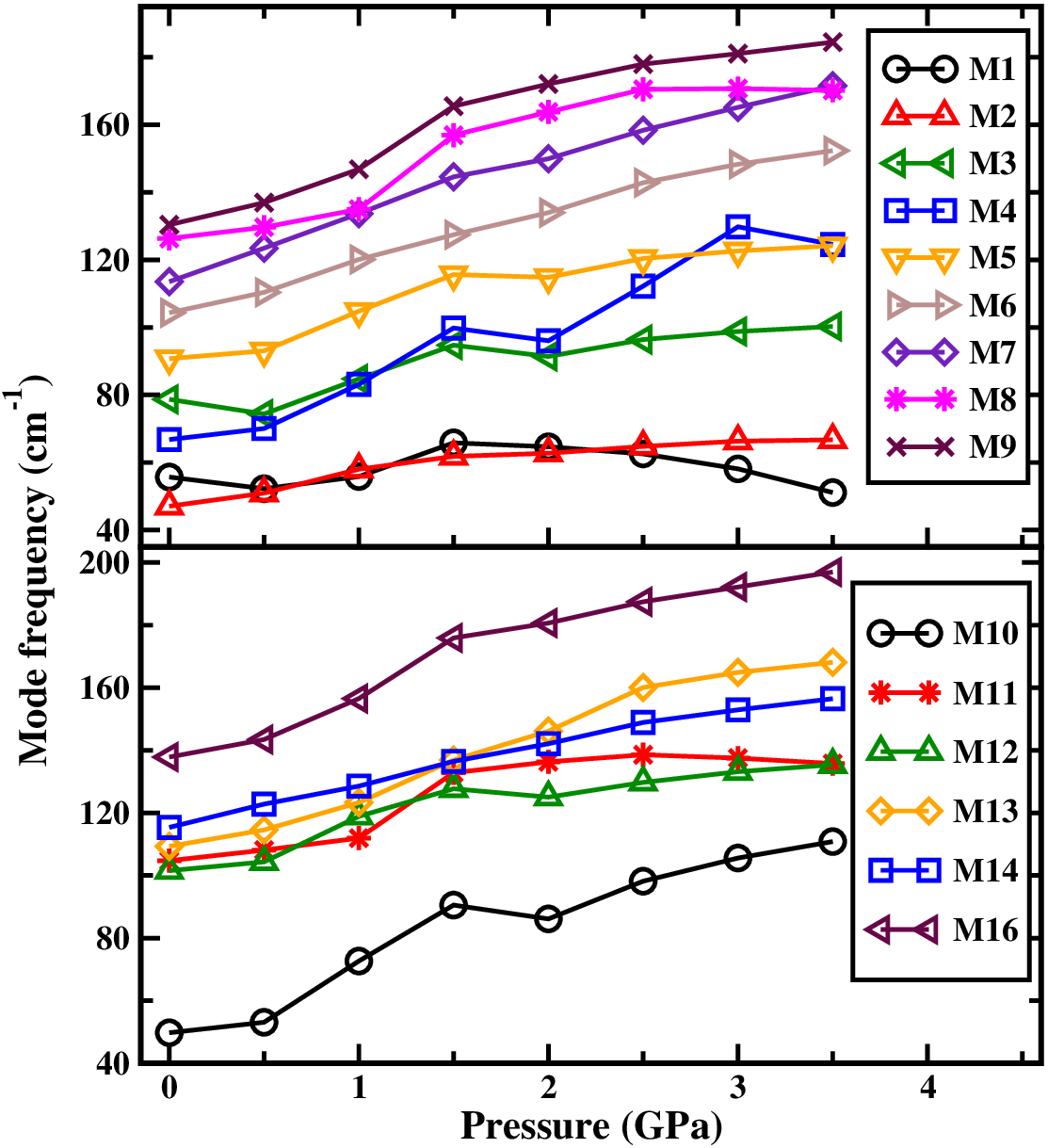} \hspace{0.05in}
\includegraphics[height = 3.5in, width=2.9in]{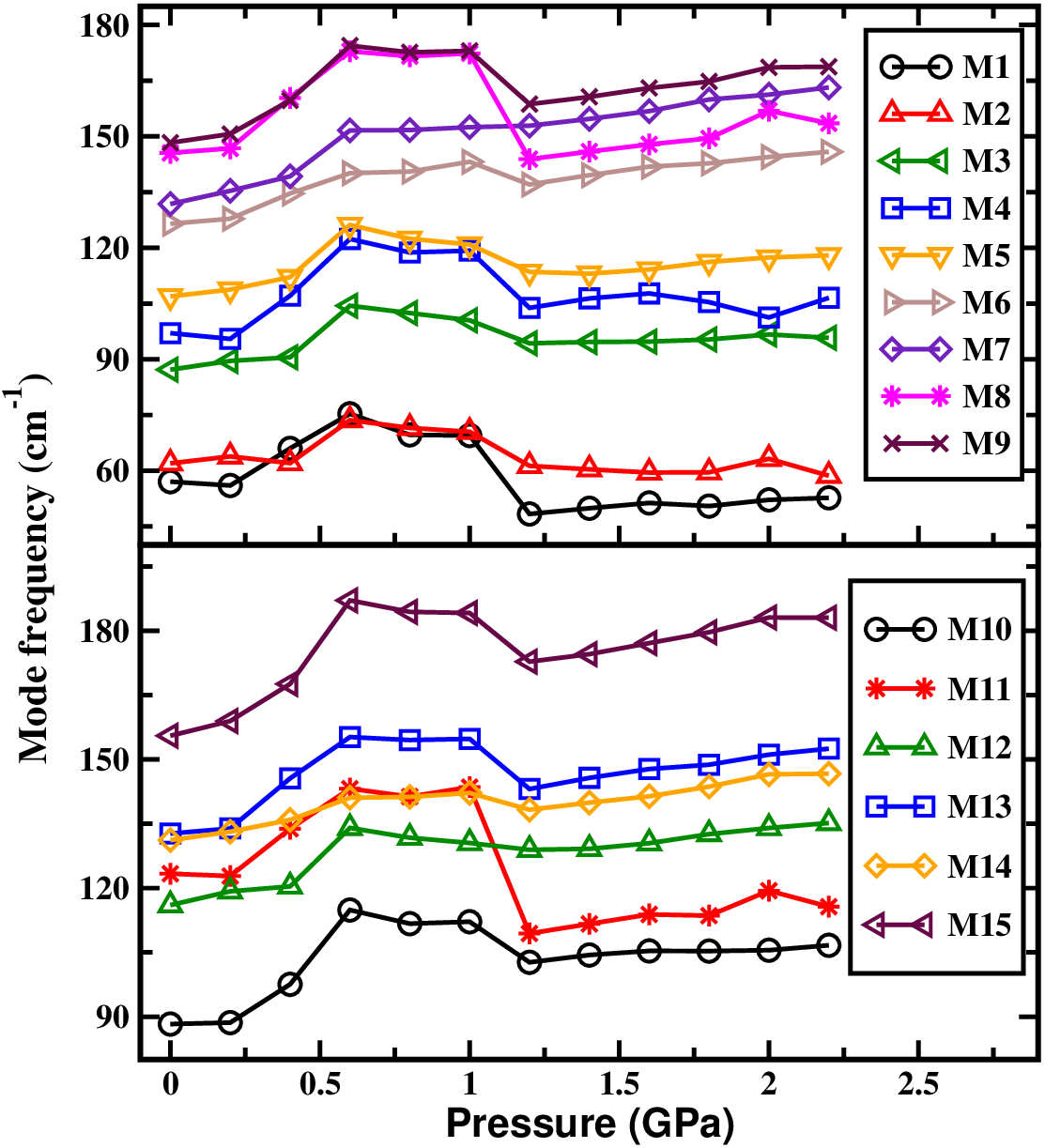}
\caption{(Colour online) Calculated a) Raman and b) IR active lattice modes of phase I of KClO$_3$ as a function of pressure using PBE (left) and PBE+G06 (right) functionals.}
\end{figure}

\begin{figure}[h]
\centering
\includegraphics[height = 3.5in, width=2.9in]{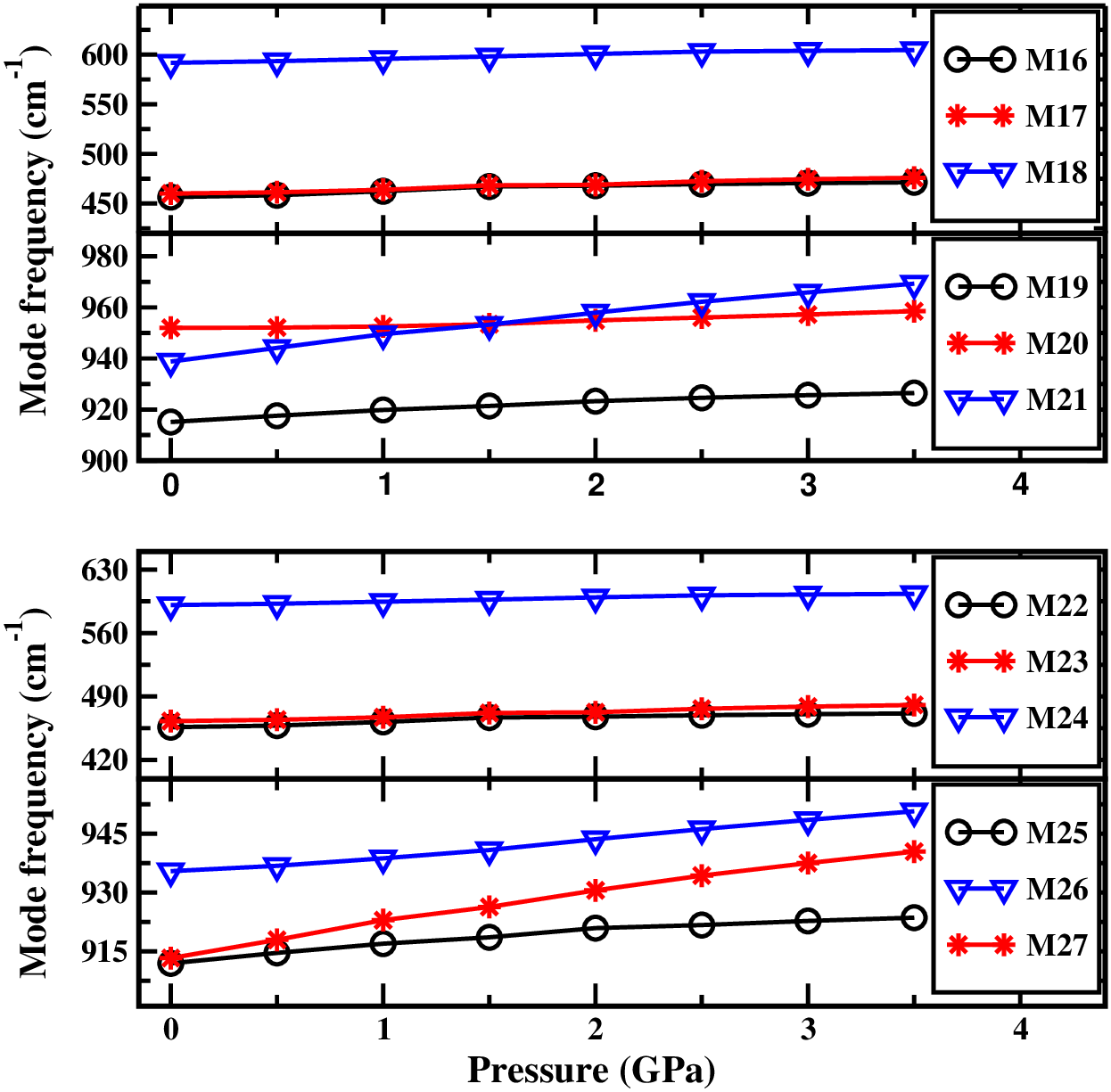} \hspace{0.3in}
\includegraphics[height = 3.5in, width=2.9in]{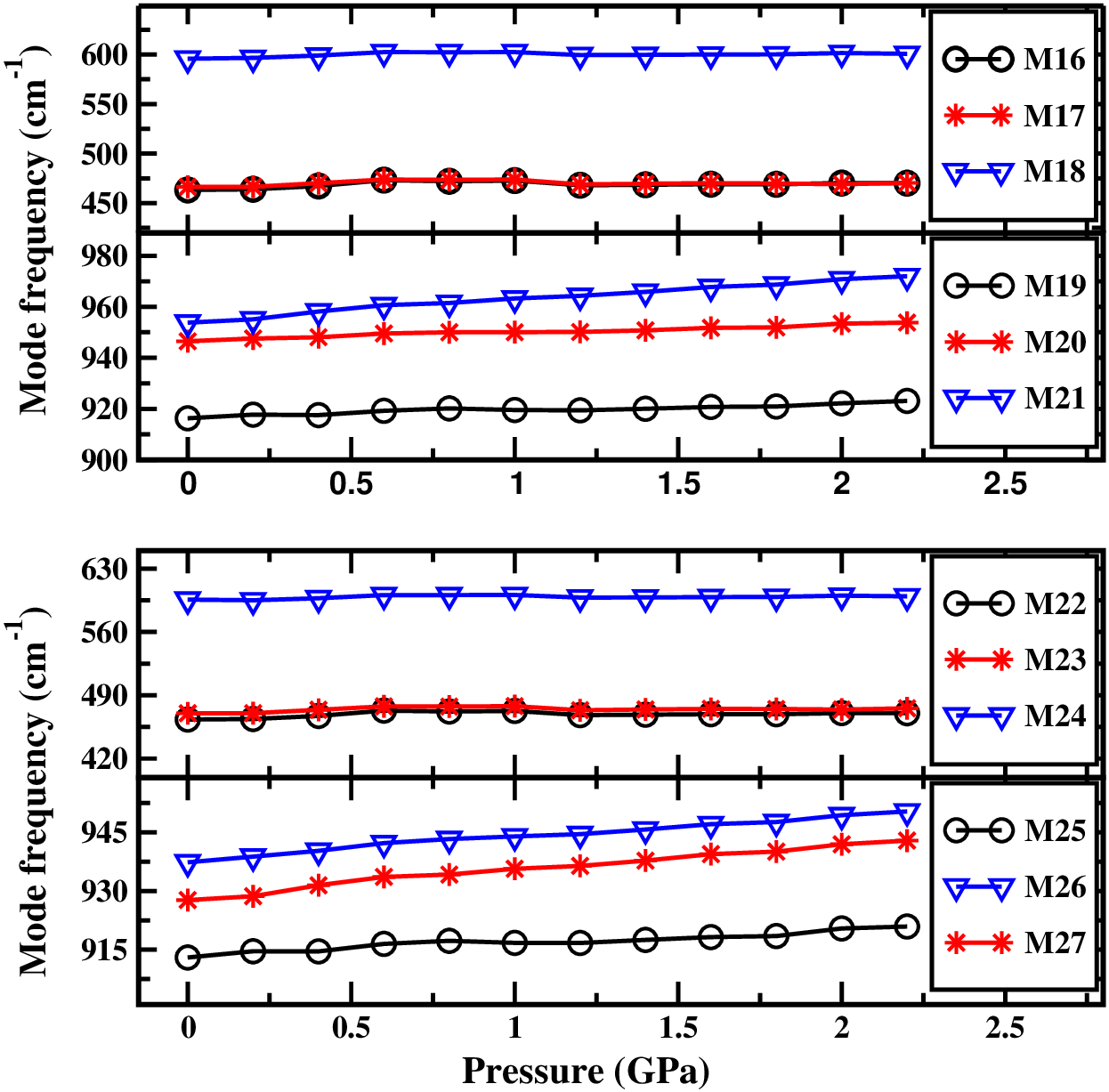}
\caption{(Colour online) Calculated a) Raman and b) IR active internal modes of phase I of KClO$_3$ as a function of pressure using PBE (left) and PBE+G06 (right) functionals.}
\end{figure}

\section{Acknowledgments}
NYK and VDG would like to thank DRDO through ACRHEM for financial support, and the CMSD, University of Hyderabad, for providing computational facilities. The authors thank Prof. C. S. Sunandana, School of Physics, University of Hyderabad, Dr. V. Kanchana, Department of Physics, Indian Institute of Technology Hyderabad for critical reading of the manuscript. Its pleasure to acknowledge Dr. T. R. Ravindran, Indira Gandhi Centre for Atomic Research, Kalpakkam for his fruitful suggestions. \\
$^*$\emph{Author for Correspondence, E-mail: gvsp@uohyd.ernet.in}

\end{document}